\newcommand{\Zl}{$Z \rightarrow l^+ l^-$}
\newcommand{\Ze}{\mbox{$Z \rightarrow e^+ e^-$}}
\newcommand{\Zmu}{\mbox{$Z \rightarrow \mu^+ \mu^-$}}
\newcommand{\Ztau}{\mbox{$Z \rightarrow \tau^+ \tau^-$}}
\newcommand{\Zlp}{$Z \rightarrow l^+ l^-$}
\newcommand{\Zep}{\mbox{$Z \rightarrow e^+ e^-$}}
\newcommand{\Zmup}{\mbox{$Z \rightarrow \mu^+ \mu^-$}}
\newcommand{\Ztaup}{\mbox{$Z \rightarrow \tau^+ \tau^-$}}
\newcommand{\Wl}{\mbox{$W \rightarrow l \nu$}}
\newcommand{\We}{\mbox{$W \rightarrow e \nu$}}
\newcommand{\Wmu}{\mbox{$W \rightarrow \mu \nu$}}
\newcommand{\Wtau}{\mbox{$W \rightarrow \tau \nu$}}

\newcommand{\spZlp}{$\sigma \times B(p \bar p \rightarrow Z \rightarrow l^+l^-)$}

\newcommand{\spZl}{$\sigma \times B(p \bar p \rightarrow Z/\gamma^* \rightarrow l^+l^-)$} 
\newcommand{\spWl}{$\sigma \times B(p \bar p \rightarrow W  \rightarrow l \nu)$}

\documentclass[epsfig,11pt,twocolumn,twoside]{article}

\usepackage[dvips]{graphics}
\usepackage{graphicx}

\begin{document}

\twocolumn[
\vspace*{-1.8cm}
\begin{flushright}
{\bf LAL 05-33}\\
\vspace*{0.1cm}
{April 2005}
\end{flushright}
\vspace*{0.8cm}

\begin{center}
{\LARGE\bf W and Z Boson Production in $p\bar p$ Collisions from\\
 Run II of the TeVatron Collider}\\
\vspace*{0.8 cm}
{\Large\bf Pierre Petroff}\\
\vspace*{0.3 cm}
{{\large\bf Laboratoire de l'Acc\'el\'erateur Lin\'eaire}\\
IN2P3-CNRS et Universit\'e de Paris-Sud, B\^at. 200 , BP 34, F-91898 Orsay}
\end{center}
\vspace*{0.8cm}

\abstract{
Measurements of inclusive W and Z cross sections times leptonic branching ratios for $p\bar p$ collisions at $\sqrt{s}$=1.96 TeV are reported here on behalf of the CDF and D\O\ collaborations. The data correspond to an integrated luminosity  up to 200 $pb^{-1}$. The ratio of leptonic W and Z rates is measured and the leptonic branching fraction B(\Wl) is extracted as well as an indirect value for the total width of the W and the CKM matrix element $|V_{cs}|$.\vspace*{0.8 cm} }]

\section{Introduction}
Testing the Standard Model (SM) is one of the main goals of the Tevatron experiments.
The large number of  W and  Z  bosons allow not only to perform a robust test of the SM but can be extensively used in the calibration of the detectors energy scale. Moreover, confidence in the results allows us to use them as normalization for successive measurements like the top cross section in which systematics and theoretical uncertainties would cancel in the ratio of the cross sections.

On behalf of the CDF and D\O\ collaborations we report new results on the W and Z cross section measurement from Run II of the Tevatron at a center mass energy of $\sqrt{s}$ = 1.96 TeV.
Interesting new result reported here is measurement of the inclusive W and Z cross sections times the branching ratio in $\tau$ \,channel. The \Ztau decay is a nice benchmark for all analyses including $\tau$'s like searches for Supersymmetry in models with high values of tan$\beta$.

\section{W and Z signatures}
Due to a large QCD background, decay channels involving quarks are difficult to measure; therefore W and Z bosons are  mainly identified through their leptonic decays.
 These decays are characterized by a high transverse energy lepton and large transverse missing energy for W, or by two high transverse energy leptons for Z.

Electrons are identified as an electromagnetic (EM) cluster using a simple cone algorithm. To reduce the background of jets faking electrons, electron candidates are required to have a large fraction of their energy deposited in the EM section of the calorimeter and pass energy isolation and shower shape requirements.
The weak bosons backgrounds and the signal have been estimated using a Monte Carlo simulation.
Electron candidates are classified as \textit{tight} if a track is matched spatially to EM cluster and if the track transverse momentum is close to the transverse energy of the EM cluster. While D\O\ results are for electrons in the central calorimeter only ($|\eta|<1.0$) CDF gives results with electrons measured in the central and plug calorimeters ($|\eta|<2.8$). Both CDF and D\O\ require at least one \textit{tight} electron in the central calorimeter ($|\eta|<1.0$) for \Ze \,candidates.

\begin{figure}[h]
\vspace*{0.3cm}
\includegraphics[width=8.cm]{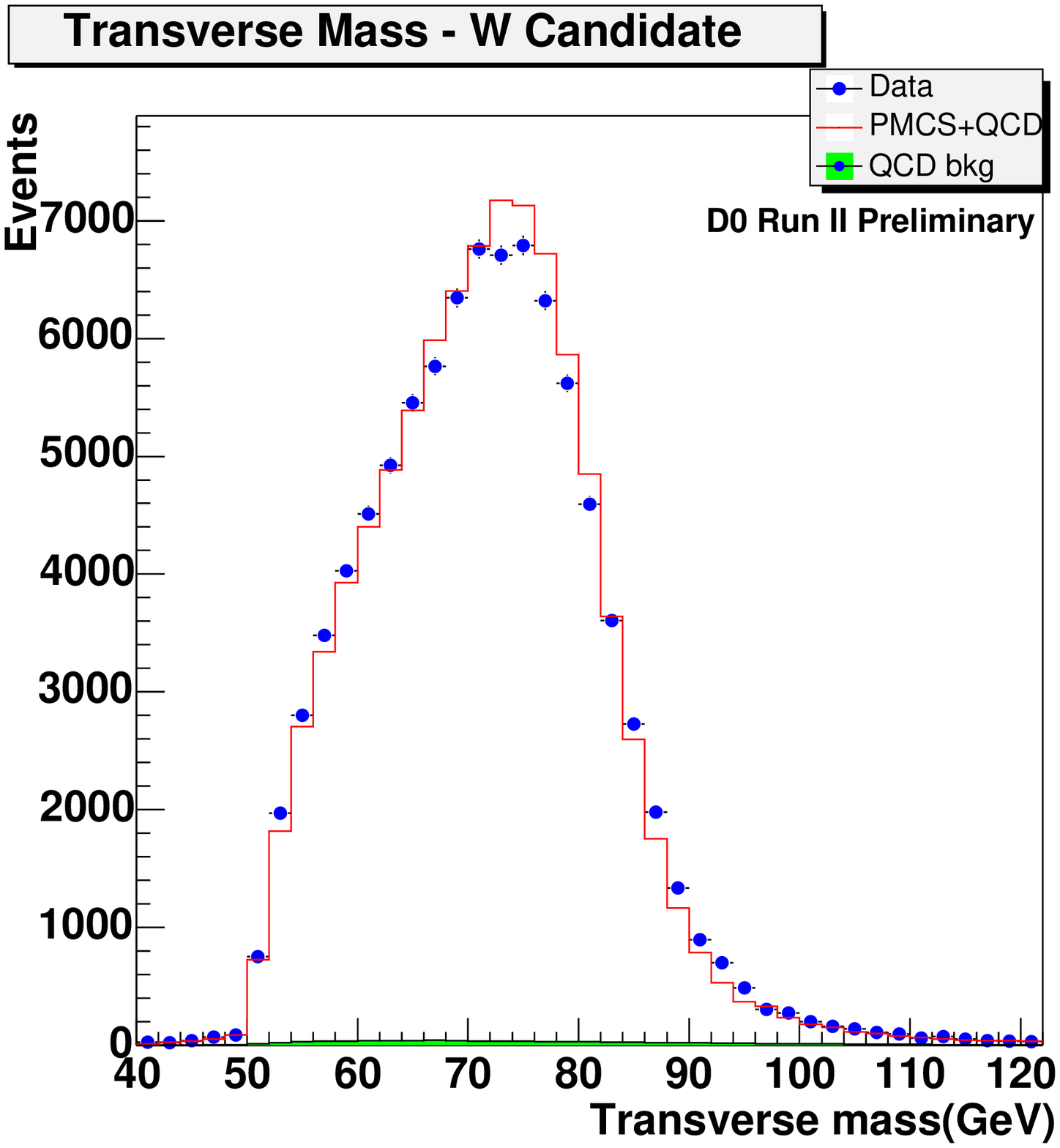}
\caption{D0: \We \,transverse mass distribution with 177 $pb^{-1}$. }
\label{fig:W-enu-mT-D0}
\end{figure}

\begin{figure}[h]
\includegraphics[width=7.5cm]{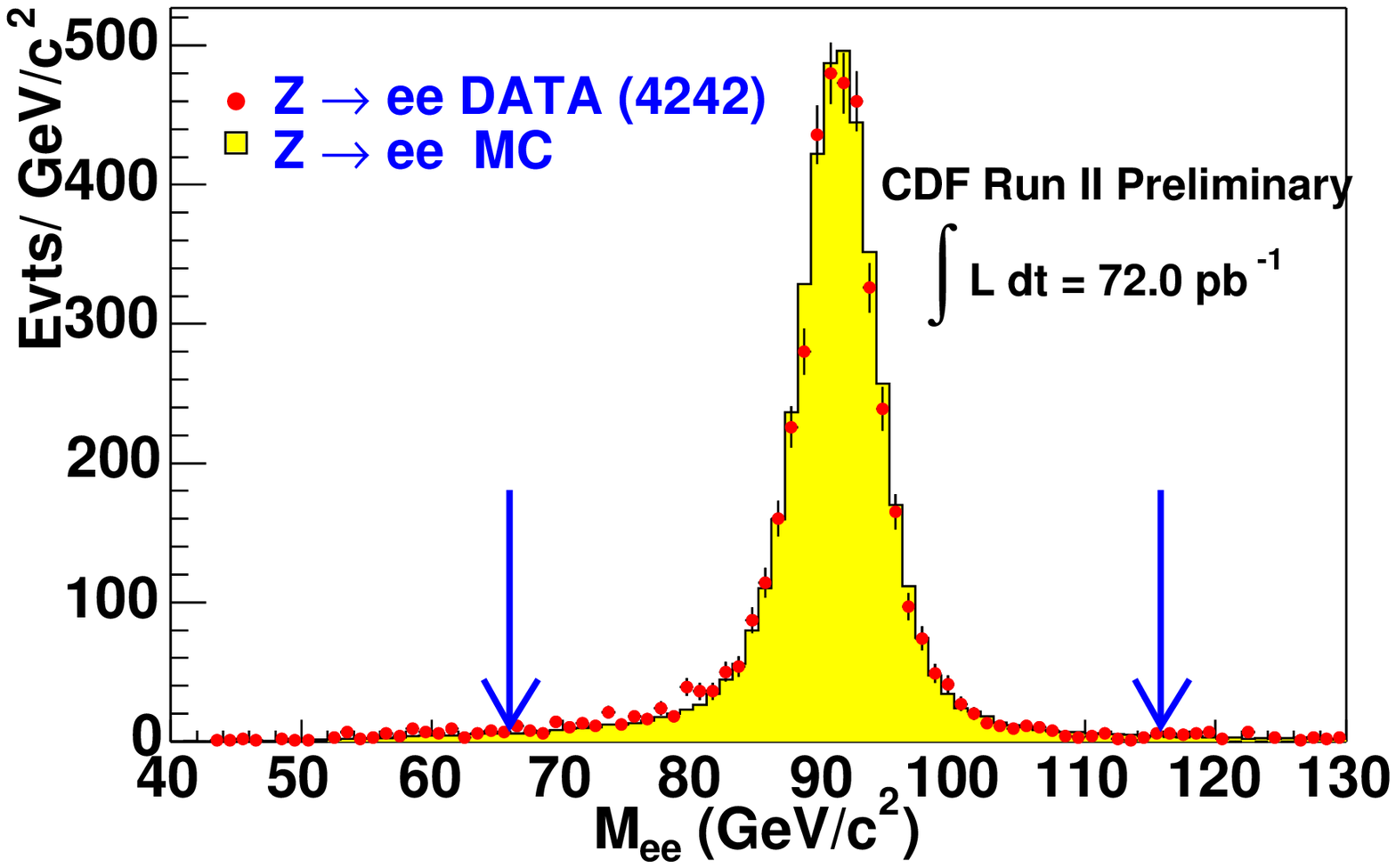}
\caption{CDF: \Ze mass distribution with 72 $pb^{-1}$. } 
\label{fig:Z-ee-CDF}
\end{figure}

The W transverse mass distribution in the channel \We\, from D\O\ is shown in Fig.~\ref{fig:W-enu-mT-D0} and the \Ze \,mass distribution from CDF is shown in Fig.~\ref{fig:Z-ee-CDF}.
These figures show that the simulation both of the background and the signal  describes the data well.

\vspace*{0.3cm}

 Muons are identified by a track in the muon system matched to a track in the central tracking system.
For CDF the measurement in the muon channel includes the muons reconstructed in the central muon extension sub-detector which extends the coverage from $|\eta|<0.6$ to $|\eta|<1.$
For D\O\ the muon reconstruction is extended to the forward muon detector with a coverage up to $|\eta| = 2.0$.

Muons from the decay of heavy-flavor hadrons are significant background to vector bosons production. It can be reduced by requiring that the muon is isolated. 
The weak bosons backgrounds  are estimated with  Monte Carlo simulation. 
Cosmic rays muons contaminate the muon sample. Timing capabilities and distance of the muon track to the vertex are used  to reduce this background to low level.

\begin{figure}[h]
\includegraphics[width=8.cm]{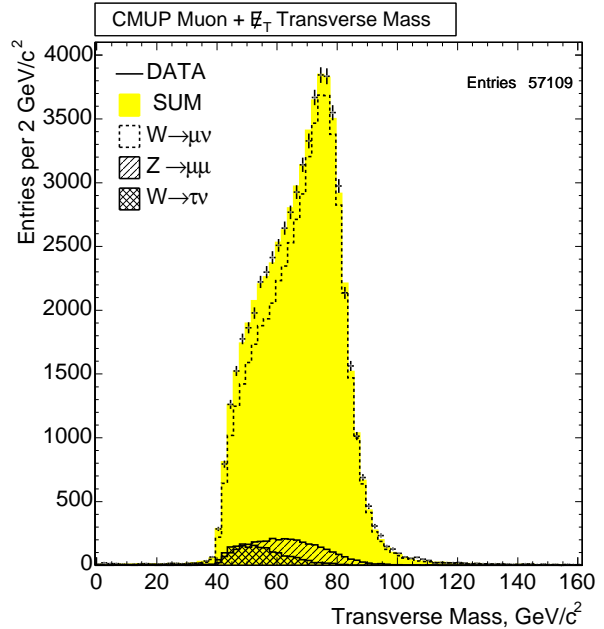}
\caption{CDF: \Wmu \ transverse mass distribution with 193.5 $pb^{-1}$. }
\label{fig:W-munu-mT-CDF}
\end{figure}

\begin{figure}[h]
\includegraphics[width=8.2cm]{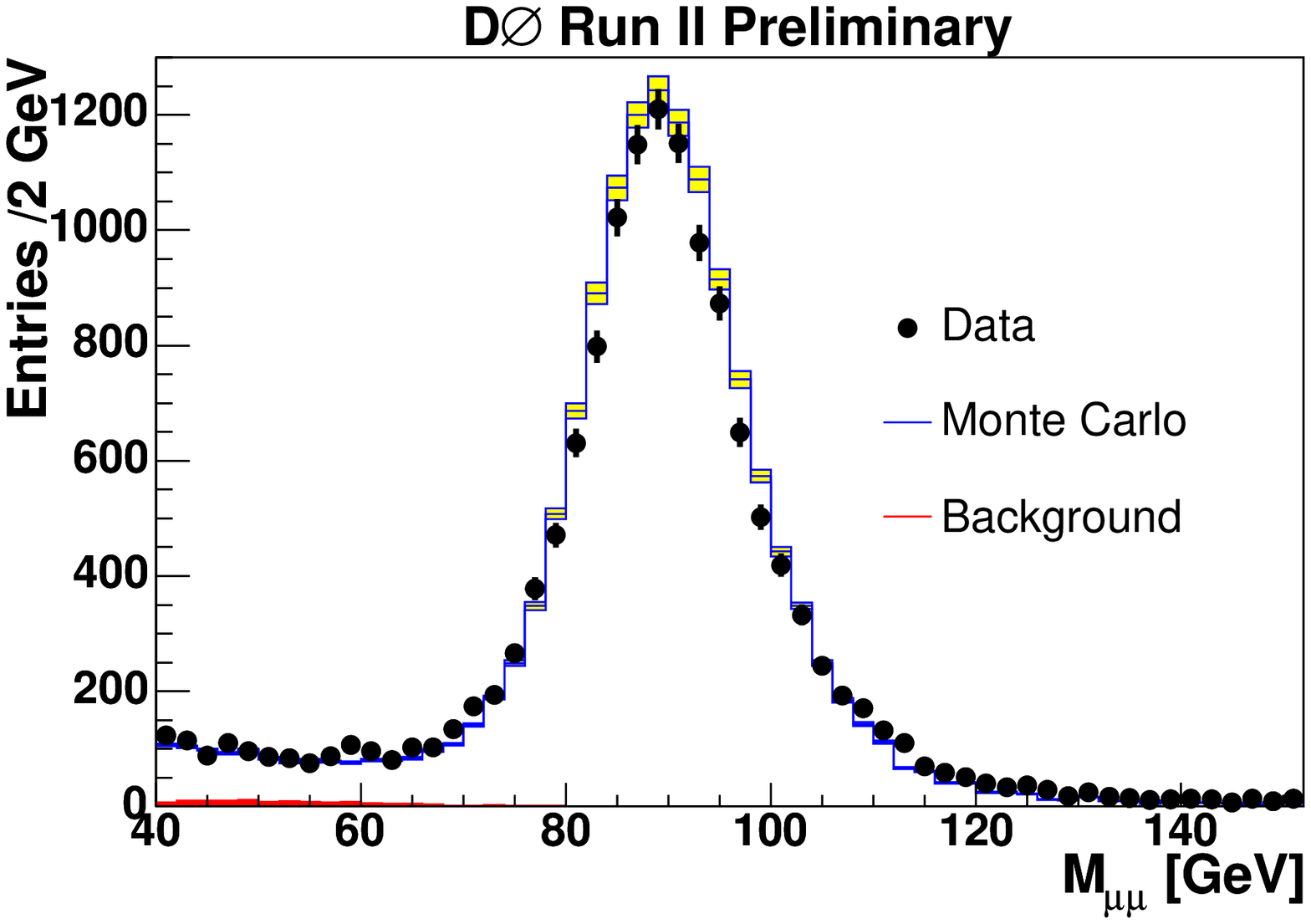}
\caption{D0: \Zmu mass distribution with 147 $pb^{-1}$. } 
\label{fig:Z-mumu-D0}
\end{figure}

The transverse \Wmu \, mass distribution in Fig.~\ref{fig:W-munu-mT-CDF} for CDF
and  \Zmu \,mass distribution for D\O\ in Fig.~\ref{fig:Z-mumu-D0}
show that backgrounds are low and the simulation is reproducing the data well. 

\begin{table*}[t]
\centering
{\normalsize
\caption{CDF: W and Z boson inclusive cross sections. \label{tab-cdf}}
\vspace*{0,5cm}
\begin {tabular}{|c|c|c|c|c|}
\hline
\raisebox{0pt}[6pt][6pt]{Channel} &
\raisebox{0pt}[6pt][6pt]{Yield} &
\raisebox{0pt}[6pt][6pt]{Bkg \%} &
\raisebox{0pt}[6pt][6pt]{${\cal L}$ $pb^{-1}$} &
\raisebox{0pt}[6pt][6pt]{$\sigma$.Br} \\
\hline
\raisebox{0pt}[6pt][6pt]{\We} &
\raisebox{0pt}[6pt][6pt]{37584} &
\raisebox{0pt}[6pt][6pt]{4.4} &
\raisebox{0pt}[6pt][6pt]{72} &
\raisebox{0pt}[6pt][6pt]{$2780\pm14_{stat}\pm60_{syst}\pm166_{lum}$ pb} \\

\raisebox{0pt}[6pt][6pt]{\Wmu } &
\raisebox{0pt}[6pt][6pt]{31722} &
\raisebox{0pt}[6pt][6pt]{9.4} &
\raisebox{0pt}[6pt][6pt]{72} &
\raisebox{0pt}[6pt][6pt]{$2768\pm16_{stat}\pm64_{syst}\pm166_{lum}$ pb } \\

\raisebox{0pt}[6pt][6pt]{\Wl } &
\raisebox{0pt}[6pt][6pt]{} &
\raisebox{0pt}[6pt][6pt]{} &
\raisebox{0pt}[6pt][6pt]{72} &
\raisebox{0pt}[6pt][6pt]{$2775\pm10_{stat}\pm53_{syst}\pm167_{lum}$ pb } \\

\raisebox{0pt}[6pt][6pt]{\Wmu } &
\raisebox{0pt}[6pt][6pt]{57109} &
\raisebox{0pt}[6pt][6pt]{9.49} &
\raisebox{0pt}[6pt][6pt]{193.5} &
\raisebox{0pt}[6pt][6pt]{$2786\pm12_{stat}{}^{+65}_{-55syst}\pm166_{lum}$ pb } \\

\hline

\raisebox{0pt}[6pt][6pt]{\Zep } &
\raisebox{0pt}[6pt][6pt]{4242} &
\raisebox{0pt}[6pt][6pt]{1.5} &
\raisebox{0pt}[6pt][6pt]{72} &
\raisebox{0pt}[6pt][6pt]{$255.8\pm3.9_{stat}\pm5.5_{syst}\pm15_{lum}$ pb } \\

\raisebox{0pt}[6pt][6pt]{\Zmup } &
\raisebox{0pt}[6pt][6pt]{1785} &
\raisebox{0pt}[6pt][6pt]{0.7} &
\raisebox{0pt}[6pt][6pt]{72} &
\raisebox{0pt}[6pt][6pt]{$248.0\pm5.9_{stat}\pm7.6_{syst}\pm15_{lum}$ pb } \\

\raisebox{0pt}[6pt][6pt]{\Zlp } &
\raisebox{0pt}[6pt][6pt]{} &
\raisebox{0pt}[6pt][6pt]{} &
\raisebox{0pt}[6pt][6pt]{72} &
\raisebox{0pt}[6pt][6pt]{$254.6\pm3.3_{stat}\pm4.6_{syst}\pm15.2_{lum}$ pb } \\

\raisebox{0pt}[6pt][6pt]{\Zmup } &
\raisebox{0pt}[6pt][6pt]{3568} &
\raisebox{0pt}[6pt][6pt]{0.4} &
\raisebox{0pt}[6pt][6pt]{193.5} &
\raisebox{0pt}[6pt][6pt]{$253.1\pm4.2_{stat}{}^{+8.3}_{-6.4syst}\pm15.2_{lum}$ pb } \\
\hline
\raisebox{0pt}[6pt][6pt]{\Wtau} &
\raisebox{0pt}[6pt][6pt]{2345} &
\raisebox{0pt}[6pt][6pt]{26} &
\raisebox{0pt}[6pt][6pt]{72} &
\raisebox{0pt}[6pt][6pt]{$2620\pm70_{stat}\pm210_{syst}\pm172_{lum}$ pb } \\
\hline
\raisebox{0pt}[6pt][6pt]{\Ztaup } &
\raisebox{0pt}[6pt][6pt]{50} &
\raisebox{0pt}[6pt][6pt]{26.5} &
\raisebox{0pt}[6pt][6pt]{72} &
\raisebox{0pt}[6pt][6pt]{$242\pm48_{stat}\pm26_{syst}\pm15_{lum}$ pb } \\
\hline
\end{tabular}
}
\end{table*}

\begin{table*}[t]
\centering
{\normalsize
\caption{D0: W and Z boson inclusive cross sections. \label{tab-d0}}
\vspace*{0,3cm}
\begin {tabular}{|c|c|c|c|c|}
\hline
\raisebox{0pt}[6pt][6pt]{Channel} &
\raisebox{0pt}[6pt][6pt]{Yield} &
\raisebox{0pt}[6pt][6pt]{Bkg \%} &
\raisebox{0pt}[6pt][6pt]{${\cal L}$ $pb^{-1}$} &
\raisebox{0pt}[6pt][6pt]{$\sigma$.Br} \\
\hline
\raisebox{0pt}[6pt][6pt]{\We} &
\raisebox{0pt}[6pt][6pt]{116569} &
\raisebox{0pt}[6pt][6pt]{3.15} &
\raisebox{0pt}[6pt][6pt]{177.3} &
\raisebox{0pt}[6pt][6pt]{$2865\pm8.3_{stat}\pm74.7_{syst}\pm186.2_{lum}$ pb} \\
\hline

\raisebox{0pt}[6pt][6pt]{\Wmu } &
\raisebox{0pt}[6pt][6pt]{8305} &
\raisebox{0pt}[6pt][6pt]{11.8} &
\raisebox{0pt}[6pt][6pt]{17} &
\raisebox{0pt}[6pt][6pt]{$3226\pm128_{stat}\pm100_{syst}\pm323_{lum}$ pb } \\
\hline

\raisebox{0pt}[6pt][6pt]{\Zep } &
\raisebox{0pt}[6pt][6pt]{4712} &
\raisebox{0pt}[6pt][6pt]{1.8} &
\raisebox{0pt}[6pt][6pt]{177.3} &
\raisebox{0pt}[6pt][6pt]{$264.9\pm3.9_{stat}\pm9.9_{syst}\pm17.2_{lum}$ pb } \\
\hline

\raisebox{0pt}[6pt][6pt]{\Zmup } &
\raisebox{0pt}[6pt][6pt]{14352} &
\raisebox{0pt}[6pt][6pt]{1.4} &
\raisebox{0pt}[6pt][6pt]{147.7} &
\raisebox{0pt}[6pt][6pt]{$291.3\pm3.0_{stat}\pm6.9_{syst}\pm18.9_{lum}$ pb } \\
\hline

\raisebox{0pt}[6pt][6pt]{\Ztaup } &
\raisebox{0pt}[6pt][6pt]{1946} &
\raisebox{0pt}[6pt][6pt]{55} &
\raisebox{0pt}[6pt][6pt]{207} &
\raisebox{0pt}[6pt][6pt]{$256\pm16_{stat}\pm17_{syst}\pm16_{lum}$ pb } \\
\hline
\end{tabular}
}
\end{table*}

The percentage of the total background for each experiment and each channel is reported in Table~\ref{tab-cdf} for CDF and in Table~\ref{tab-d0} for D\O\ . In these tables
 the total number of events and the corresponding total integrated luminosity are quoted also.

The $\tau$ lepton reconstruction is challenging at hadron collider.
CDF has performed a cross section measurement of the channel \Wtau. One or three charged tracks with $\pi^0$'s are selected in a $10^0$ cone pointing toward a narrow calorimeter cluster. The combined mass of tracks and $\pi^0$'s should be less than 1.8 GeV/$c^2$. The main background is coming from jets and weak bosons leptonic decays. 
The channel \Ztau is studied by CDF in a mode with one $\tau$ decaying leptonically in an electron and the other   decaying into hadrons. The $\tau$ candidates are reconstructed by matching narrow calorimeter clusters with tracks following the same method used in the \Wtau \, analysis. 


The number of events, the background and the integrated luminosity for the CDF  analysis can be found in Table~\ref{tab-cdf}.

\begin{figure}[h]
\includegraphics[width=7.8cm]{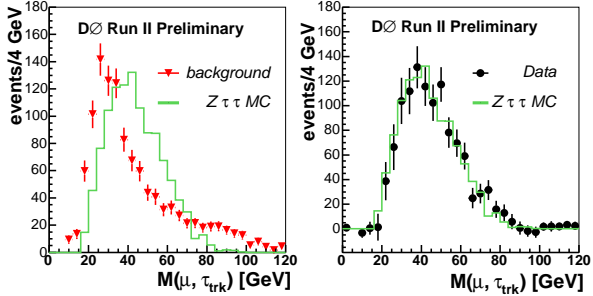}
\caption{D0: \Ztau invariant mass distribution of the $\mu$ and the $\tau$ tracks for background and data overlayed on \Ztau normalized MC.} 
\label{fig:Z-tautau-mutrk-D0}
\end{figure}

D\O\ has performed a measurement in the channel \Ztau using a Neural Network techniques.
A $\tau$ decays into $\mu \nu_{\mu} \nu_{\tau}$ and the other into hadrons+$\nu_{\tau}$ or $e\nu_e\nu_{\tau}$. The total background, the number of events and the integrated luminosity are given in Table~\ref{tab-d0}.
Fig.~\ref{fig:Z-tautau-mutrk-D0} shows the invariant mass distribution of the  $\mu$ and the $\tau$ tracks both for background estimated from like sign (LS) sample and \Ztau signal from opposite sign (OS) sample.

\section {W and Z cross sections}

\begin{figure}[h]
\includegraphics[width=7.6cm]{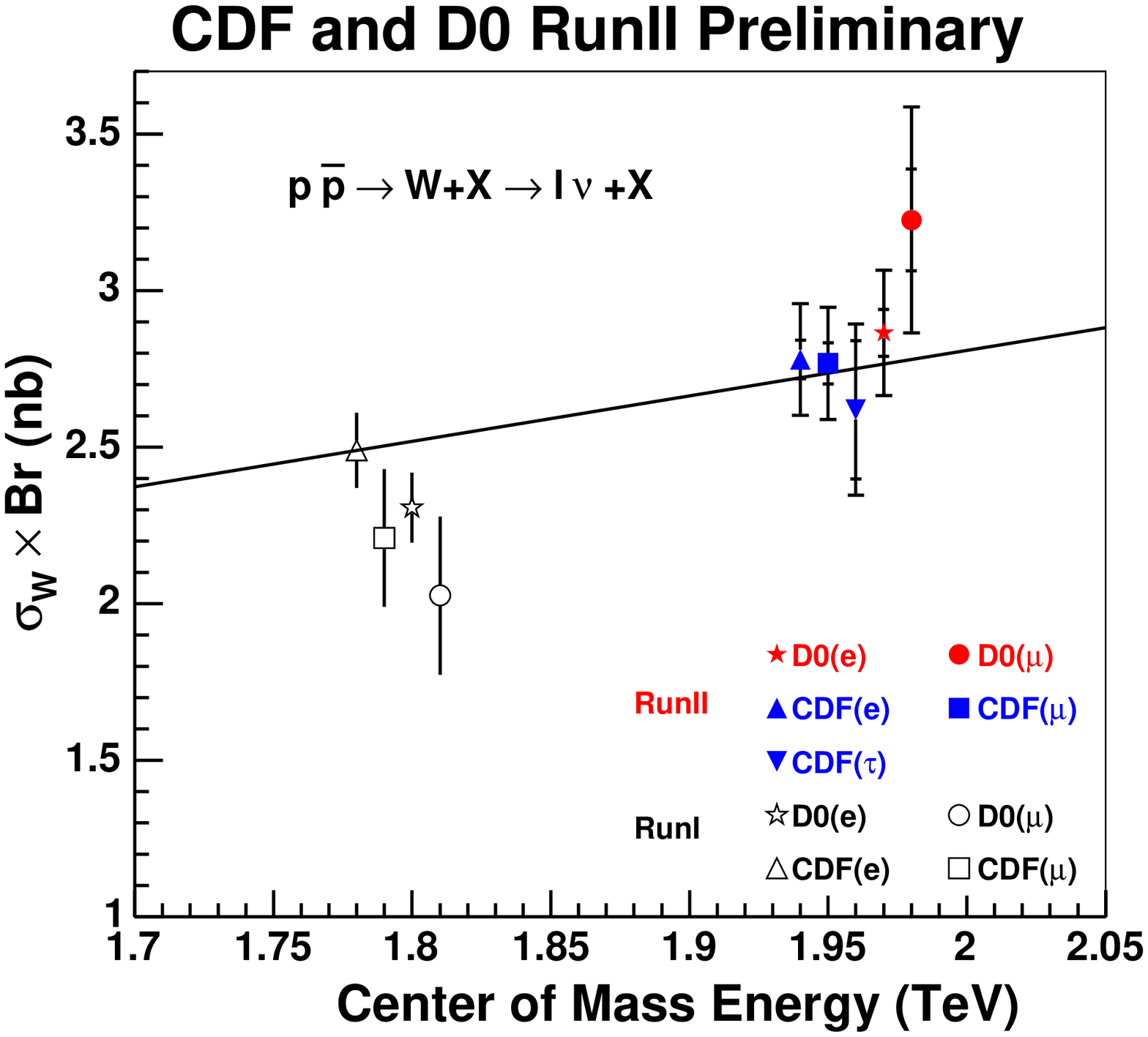}
\caption{\spWl \,cross section measured by CDF and D0 and theoretical predictions.}
\label{fig:CDF-D0-W-lnu}
\vspace*{0.5cm}
\includegraphics[width=7.6cm]{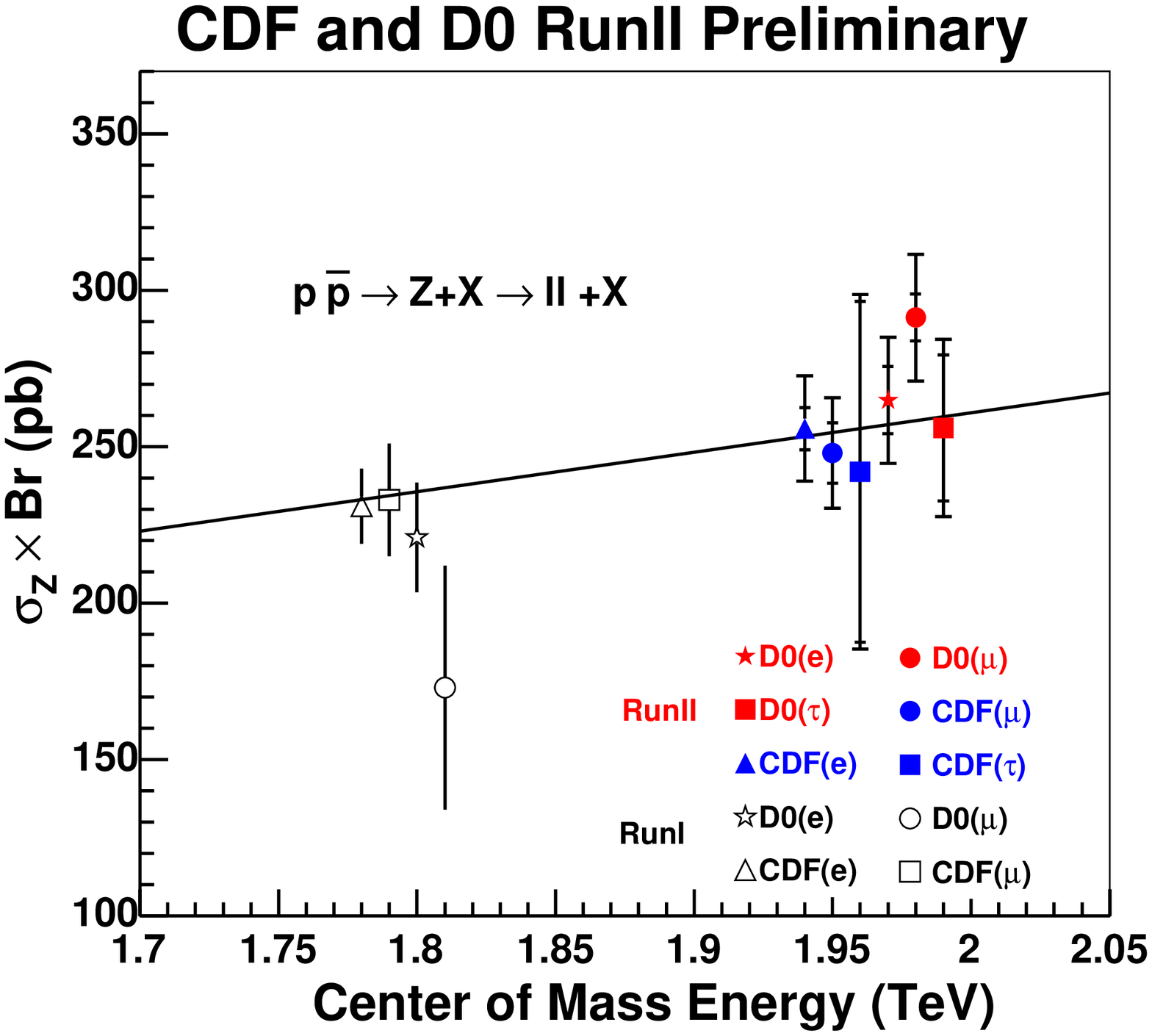}
\caption{\spZlp \,cross sections measured by CDF and D0 and theoretical prediction.} 
\label{fig:CDF-D0-Z-ll}
\end{figure}

The cross section measurements are listed in Table~\ref{tab-cdf} and Table~\ref{tab-d0} for CDF \cite{cdf} and D\O\ respectively after subtraction  of the virtual photon exchange in the \spZl \,channels.  Combined cross section measurements for \textit{e} and $\mu$ channels at ${\cal L}$ = 72 $pb^{-1}$ are reported in Table~\ref{tab-cdf} by CDF.
Fig.~\ref{fig:CDF-D0-W-lnu} and Fig.~\ref{fig:CDF-D0-Z-ll} show a good agreement with NNLO theoretical calculation \cite{th}. 

For systematic uncertainties
the main contribution  $\sim 6 \%$ is coming from the luminosity measurement. The PDF uncertainties are $\sim 2 \%$ and are estimated using the eigenvector basis sets for CTEQ6M \cite{pdf}. The contribution from the lepton  identification is $\sim 2\%$. This error is expected to be reduced with more statistics in the future.

The largest uncertainty on the cross section measurements comes from the uncertainty on the luminosity measurements. This uncertainty cancels when determining the ratio of the W boson to Z boson production cross section.
\newpage

CDF reports results on the ratio R of the leptonic W and Z cross sections for combined \textit{e} and $\mu$ channels and for the $\mu$ channel only with a sensitivity of 72 $pb^{-1}$ and 193.5 $pb^{-1}$ respectively. 
R = $10.92\pm0.15_{stat}\pm0.14_{syst}$ for the combined channels and R = $11.02\pm0.18_{stat}{}^{+0.17}_{-0.14syst}$ for the $\mu$ channel.
D\O\ reports a measurement of R = $10.82\pm0.16_{stat}\pm0.28_{syst}$ in the \textit{e} channel.

Using the measured value B(\Zl)\footnote{B(\Zl) = $0.033658\pm0.000023$} at LEP and a theoretical calculation of the ratio of production cross sections \cite{thr}, CDF extracts the leptonic branching ratio B(\Wl) = $10.89\pm0.2$ \% for the combined channels  and  B(\Wmu) = $11.01\pm0.18_{stat}{}^{+0.18}_{-0.15syst}$ \% for the $\mu$ channel. 
From the theoretical value of the leptonic partial width, $\Gamma$(\Wl)\footnote{$\Gamma(\Wl) = 226.4\pm0.3$ MeV} \cite{wl}, CDF extracts the total width of the W boson: $\Gamma^{tot}_W = 2079\pm41$ MeV. This values is close to the SM value and the current world average value \footnote{ Current world average = $\Gamma^{tot}_W = 2118\pm42$ MeV and SM value = $2092.1\pm2.5$ MeV}.
Finally as in the SM the total $\Gamma^{tot}_W$ width depends on certain CKM matrix elements
CDF determined the matrix element $|V_{cs}| = 0.967\pm0.03$ by fixing all other matrix elements at their world average values.

\section{Conclusion and Prospects}
W and Z production cross section values have been measured by CDF and D\O\, showing an agreement with SM predictions. The increase of the statistics and the combination of the results of the two experiments will allow in a  near future to improve the accuracy of the test of the Standard Model.  
Finally the first results on the \Wtau \, and \Ztau \, cross sections demonstrate the feasibility of a new physics search with $\tau$'s at the Tevatron.

\end{document}